\documentclass[10pt]{article}
\usepackage{microtype}
\usepackage{graphicx}
\usepackage{geometry}   %
\usepackage{subfigure}
\usepackage{booktabs} %
\usepackage[pagebackref]{hyperref}       %
\usepackage{tabularx}
\hypersetup{
  colorlinks=true,
  linkcolor=blue,
  filecolor=magenta,
  urlcolor=cyan,
  citecolor=blue
}
 \renewcommand*{\backrefalt}[4]{%
     \ifcase #1 \footnotesize{(Not cited).}%
     \or        \footnotesize{(Cited on page~#2).}%
     \else      \footnotesize{(Cited on pages~#2).}%
     \fi}

\usepackage{hh}
\usepackage{authblk}
\RequirePackage{fancyhdr}
\RequirePackage{xcolor} %
\RequirePackage{algorithm}
\RequirePackage{algorithmic}
\RequirePackage{natbib}
\RequirePackage{eso-pic} %
\RequirePackage{forloop}
\setcitestyle{authoryear,round,citesep={;},aysep={,},yysep={;}}
\usepackage{amsmath}
\usepackage{amssymb}
\usepackage{mathtools}
\usepackage{amsthm}
\usepackage{thmtools}
\usepackage{thm-restate}

\declaretheorem[numberwithin=section]{definition, assumption, remark}
\usepackage[noabbrev]{cleveref}
\crefname{assumption}{assumption}{assumptions}
\usepackage[disable,textsize=tiny]{todonotes}

\title{\bf Mobility-based Traffic Forecasting in a Multimodal Transport System}
\author[1]{Henock M. Mboko}
\author[2]{Mouhamadou A.M.T. Bald\'e}
\author[2]{Babacar M. Ndiaye}

\affil[1]{African Institute for Mathematical Sciences (AIMS), Mbour, Senegal. {\it henock.s.makumbu@aims-senegal.org}}
\affil[2]{Laboratory of Mathematics of Decision and Numerical Analysis, 
Cheikh Anta Diop University. Po Box 45087, 10700. Dakar, Senegal.  {\it mouhamadoualioumountagatall.balde\_babacarm.ndiaye@ucad.edu.sn}}

\date{}

\begin{document}

\maketitle

\begin{abstract}
We study the analysis of all the movements of the population on the basis of their mobility from one node to another, to observe, measure, and predict the impact of traffic according to this mobility. 
The frequency of congestion on roads directly or indirectly impacts our economic or social welfare. Our work focuses on exploring some machine learning methods to predict (with a certain probability) traffic in a multimodal transportation network from population mobility data. We analyze the observation of the influence of people’s movements on the transportation network and make a likely prediction of congestion on the network based on this observation (historical basis).\\


\end{abstract}

%---%
\section{Introduction}
The government of Senegal had expressed the importance to  transport sector and urban mobility. The vision in transport sector explains the importance of the investments planned in the emerging senegalese plan. In particular, the priority action plan, in relation to urban mobility: the etablishing of the Regional Express Train (TER), the Bus Rapid Transit (BRT), and the reinforcement of the Dakar Dem Dikk fleet (400 city buses out of 475 vehicles) \citep{cetudweb}. In addition, we have the continued renewal of the urban fleet of fast buses and ndiaga ndiaye (more than 60\%), together with the professionalization of informal or artisanal actors. \\
Urban transport malfunctions (congestion, pollution, and accidents) are recorded in some African countries at about 4\% of GDP. The movement of an individual in a network is simply a succession of nodes and arcs. For a given individual, let us assume that we only know his path to a certain node. \\
%context
\indent Many urban areas in the world, especially in developing countries, are facing a rapid increase in population density, that generates a transport demand that cannot be supported by transport infrastructures. 
Between 1976 and 2022, the population of the Dakar region increased approximately 67 times while at the same time the transport network and urban planning were not sufficiently adapted to this development \citep{ansd}. This leads to congestion problems and a reduction in urban accessibility defined as the ability to access certain given resources or activities, within a given time frame \citep{gueye}.\\
%Problems and Objectives
\indent Is it possible, by analyzing all movements, to predict (with a certain probability) the next node where this individual will go? The scientific fields associated with this question are related to pattern recognition and machine learning. These two fields make extensive use of statistical and operations research techniques. The underlying application is traffic forecasting. This new research direction is based on the idea that there are recurrent trajectory patterns hidden in CDR files, the knowledge of which allows to reduce considerably the amount of files needed. And these patterns can be discovered by Machine Learning on the data corpus (and others). Being able to
make such predictions gives a definite advantage on the operational management. \\
%Topic Originality
\indent This work is positioned on resolution approaches. There are currently few methods, especially exact, to solve this type of problem by considering the system in its globality. Knowing 
where individuals start from (origins), where they go (destinations), what they do (activities), which paths they follow, and by which means of transport (mode) is essential information whose real-time knowledge can inform urban management policies. We aim at, is a system that exploits in real-time this kind of files in. However, such an objective poses many computational and mathematical challenges.\\
- {\bf Challenge 1}: Exploration of data mining methods. First of all, the data analysis proposed by Gueye in \citep{gueye}, Baldé in \citep{balde1, balde2}, Kone in \citep{kone1, kone2}  consisted of a series of algorithms based on simple and improvable heuristic rules. 
However, the extraction of knowledge from data is precisely the domain of data mining and multi-agent systems, using statistical techniques. We have not exploited all the possibilities. One of this research directions will be to explore this path.\\
- {\bf Challenge 2}: Traffic prediction by machine learning. Secondly, having real-time information based on the exploitation of data files must take into account the momentary absence of these 
files. It is indeed difficult to envisage, because of the construction delays and their sizes, to have a system regularly fed, on very short time steps by data. And capable, on equally short time steps, of extracting the relevant information. We have also experimentally observed, in the challenge, that not all trajectories are relevant to study. Indeed, some of them contain too many
information gaps to be exploitable because the individuals are only detected if they receive or send a call/sms or by Google if they have turned on the Location History setting, Google Mobility Data \citep{aktay}. And the trajectories can be grouped into clusters (of trajectories) sharing approximately the same characteristics: almost the same origins, same destinations, same
activities, same modes. Only one trajectory among each cluster is then needed to report on all of them.\\
\indent For example, let us assimilate the transportation network to a graph whose arcs represent the roads and nodes their intersections. The movement of an individual in this network is just a succession of nodes and arcs taken. For a given individual given, let us admit that we only know his path to a certain node. Is it possible, by analyzing all the trips, to predict (with a certain probability) the next node where this individual will go? \\
\noindent The article is organized as follows. In Section \ref{prophetsec}, we present the mathematical model, error measurement and some advantages for the Prophet forecasting. In Section \ref{numresults}, we numerically determine the Dakar traffic forecast by analyzing the mobility time series and improving the model prediction. Finally, in Section \ref{ccl}, we present conclusions and perspectives.

%---%
\section{The Prophet forecasting model}\label{prophetsec}
\subsection{Mathematical model}
Prophet is a time series model that may be decomposed into three components: trend, seasonality, and holidays. They are put together in the following equation: 
$$
y(t) = g(t) + s(t) + h(t) +\epsilon
$$
with $g(t)$ : trend function, $s(t)$ : seasonalities, $h(t)$ : holidays, $\epsilon_t$ : is a white noise error term. \\
This specification is akin to a generalized additive model (GAM) \citep{hastie, milion}, a type of regression model that includes possibly non-linear smoothers on the regressors. In this case, the time is used as a regressor, but many linear and nonlinear time functions are included as components. 
Seasonality is modeled as an additive component in the same way that exponential smoothing \citep{gardner}. A log transform can be used to provide multiplicative seasonality, where the seasonal impact is a factor that multiplies $g(t)$.
\begin{enumerate}
\item {\bf The trend model} :\\
The trend function models non-periodic changes in the values of the time series. Two trends models are implemented : saturating growth model and piecewise linear model.
 \begin{itemize}
 \item {\bf Saturating growth model} \\
 The frame of growth can be explained on the basis of the evolution or increase of a series of values. The central element of the data generation process is a model that explains how, for example, the population has increased and how it should continue to increase. Facebook's growth modeling is often similar to the population growth in natural ecosystems, where there is nonlinear growth that saturates at a carrying capacity. For example, the load capacity for the number of Facebook users in a particular area could be the number of people with access to the Internet. In the same way, the carrying capacity of the displacement or circulation of the population in Dakar can also be explained in relation to the number of people who have activated the localization via Google Map. This modeling can reflect the image of the state of people's mobility according to certain strategic zones in the study of possible traffic congestion. Saturation can express a high concentration of population in an area, which is interpreted by the
carrying capacity \citep{taylor}. The logistic growth model, in its most basic version, is used to represent this type of growth :
\begin{equation}
g(t) = \frac{C}{1 + exp(-k(t - m))}
\end{equation}
with $C$ the carrying capacity, $k$ the growth rate, and $m$ an offset parameter. 
Taking into account two important aspects of population mobility growth: carrying capacity and variation in growth rate, as well as considering trend changes in the growth model by explicitly designating changepoints where the growth rate is permitted to change.
Assume that there are $S$ changepoints at periods $s_j$, $j = 1,...,S$. We create a rate adjustment vector $\delta \in \mathbb{R}^S$,  where $\delta_j$ is the rate change that happens at time 
$s_j$. The rate at any moment t is thus the base rate $k$ plus any $P$ modifications up to that point: $k + \sum_{j:t>s_j} \delta_j$.   This is more clearly stated by creating a vector $a(t) \in [0, 1]^S$, such that
$$
a_j(t) = \left\{\begin{array}{cl}
  1 &  t \geq s_j \\
  0 &  \mbox{otherwise} \\
\end{array}\right.
$$
The rate at time t is then given by $k + a(t)^T\delta$. When the rate $k$ is changed, the offset parameter $m$ must also be changed to link the segment ends. The right adjustment at changepoint $j$ may be calculated easily as 
\begin{equation}
\gamma_j = \Big(s_j - m - \sum_{l<j} \gamma_l\Big) \Big(1- \frac{k+\sum_{l<j} \delta_l}{k+\sum_{l<j} \delta_l}  \Big)
\end{equation}
The piecewise logistic growth model is then
\begin{equation}
g(t) = \frac{c(t)}{1 + exp(-(k + a(t)^T\delta)(t - (m + a(t)^T\gamma)))}
\end{equation}
$C(t)$, or the predicted capabilities of the system at any point in time, is an essential set of parameters in this model. Analysts frequently have knowledge of market sizes and may adjust them properly. External data sources, such as World Bank population predictions, may also be available to provide carrying capacities. The logistic growth model provided here is a subset of generalized logistic growth curves, which are all sigmoid curves. It is simple to extend this trend model to different families of curves \citep{taylor}.

\item {\bf Linear trend with changepoints}\\
A piecewise constant rate of growth provides a simple and frequently helpful model for predicting issues that do not display saturation growth. Here The trend model is used :
\begin{equation}
g(t) = (k + a(t)^T\delta)(t + (m + a(t)^T\gamma)
\end{equation}
$k$ is the growth rate, $\delta$ has the rate adjustments, $m$ is the offset parameter, and $\gamma_j$ is set to $-s_j\delta_j$ to make the function continuous \citep{taylor}.
 \end{itemize}
\item  {\bf Seasonality} :\\
Time series, such as for our case of mobility of people often represent a multi periodicity depending on the behavior of humans. To adjust and predict these effects, in the Prophet model, seasonality models have been specified which are periodic functions of $t$, relying on Fourier series to provide a flexible model of the effects periodicals \citep{harvey}. Let $P$ be the regular period (for example, $P$ = 365.25 for annual data or $P$ = 7 for weekly data,  when we measure our time variable in days).  
It is approximated arbitrary smooth seasonal $N$-effects with 
\begin{equation}
s(t) = \sum_{n=1}^N \Big(a_n \cos\Big(\frac{2\pi nt}{P})+ b_n \sin\Big(\frac{2\pi nt}{P}\Big)\Big)
\end{equation}
a standard Fourier series 2. Seasonality must be fitted by estimating the $2N$ parameters $\beta=[a_1, b_1,..., a_N, b_N]^T$ . This is done by constructing a matrix of seasonality vectors for each value of t in the historical and future data, for example with yearly seasonality and $N$ = 10.
\begin{equation}
X(t) = \Big[\cos\Big(\frac{2\pi(1)t}{365.25}\Big),..., \sin\Big(\frac{2\pi(10)t}{365.25}\Big)\Big]
\end{equation}
The seasonal component is then
\begin{equation}
s(t) = X(t)\beta
\end{equation}
In the generative model, we take $\beta \sim Normal(0,\sigma^2)$ to impose a smoothing prior on the seasonality.

\item  {\bf Holidays and events} :\\
Holidays and events provide large and somewhat predictable shocks to many business
time series and often do not follow a periodic pattern, so their effects are not well modeled by a smooth cycle. Prophet models them, providing an opportunity to create a data frame for them. It has two columns (holiday and ds) and one row for each occurrence of the holiday. It should include all occurrences of the holiday, both in the
past (as far back as historical data goes) and in the future (as far back as forecasts are made). If they do not occur in the future, Prophet will model them and not include them in the forecast.
By assuming that the impacts of holidays are independent, it is simple to incorporate this list of holidays into the model. Allow Di to be the collection of previous and future dates for each holiday i. We include an indicator function that indicates if time has passed.\\
$t$ is during holiday $i$, and assign each holiday a parameter $k_i$ which is the corresponding change in the forecast. This is done in a similar way as seasonality by generating a matrix of regressors
\begin{equation}
Z(t) = [1(t \in D_1),...,1(t \in D_L)]
\end{equation}
and taking
\begin{equation}
h(t) = Z(t)k
\end{equation}
As with seasonality, we use a prior
\begin{equation}
k \sim Normal(0, \sigma^2)
\end{equation}
\end{enumerate}

\subsection{Error measurement}
Prophet has a time series cross-validation capability that uses past data to calculate forecast inaccuracy. This is accomplished by identifying cut-off points in the history and fitting the model with data only up to that point for each one. The projected values can then be compared to the actual values.
The cross-validation function can automatically perform this cross-validation technique for various historical cutoffs. We define the forecast horizon (horizon), as well as the size of the first learning period (initial) and the spacing between cutoff dates (period). The first learning period is set to three times the horizon by default, with cuts every half-horizon. 
Cross-validation produces a data frame containing the actual y-values and the out-of-sample forecast values $yhat$ for each simulated forecast date and cutoff date. A prediction is generated for each observed point between the cutoff date and the cutoff date plus
horizon. This data frame can then be used to calculate the $yhat$ vs. $y$ errors.

\subsection{Advantages of model}
The GAM formulation has the benefit of simply decomposing and accommodating new components as needed, such as when a new source of seasonality is detected. GAMs also fit relatively fast, either via backfitting or L-BFGS \citep{byrd}, allowing the user to alter the model parameters interactively.
While we give up some crucial inferential benefits of employing a generative model such as ARIMA, this formulation has a number of practical advantages:
 \begin{itemize}
 \item Flexibility: With various periods, we can easily handle seasonality and allow the analyst to make varied assumptions about trends.
\item  Unlike ARIMA models, data does not need to be consistently spaced, and missing values, such as those resulting from outlier removal, do not need to be interpolated.
\item Fitting is extremely fast, allowing the analyst to interact with several model specifications, such as in a Shiny application \citep{chang}.
\item  The forecasting model includes clearly interpretable parameters that analysts may adjust to impose assumptions on the projection. Furthermore, analysts often have regression experience and can readily modify the model to accommodate additional components.
 \end{itemize}

More information about Prophet can be found in \citep{prophet}.

%---%
\section{Numerical simulations}\label{numresults}
\subsection{Data source}
At some point, people's actions were influenced by the covid-19 pandemic with regard to displacement, which is a mobile habit. And today's society is struggling to regulate this mobility, as it has an impact on the local economy driven by various daily activities often facilitated by the means of travel. To understand the evolution of mobility, the government needs a road map. The information gained can be used as a guide to make informed decisions. These choices are made to mitigate the hazards and impacts that the community will experience. \\
In this work, we are interested in the transport sector by forecasting traffic in relation to the mobility of individuals in Dakar city, and visualizing the overall situation in some regions of Senegal.\\
We used mobility data from Google \citep{aktay}. It should be noted that Google began publishing the first version of the community mobility reports in early April 2020 with the goal of providing global insights that can help policymakers combat covid-19. \\
These reports plot movement patterns over time by geography in different categories of locations such as retail and leisure, grocery and drug stores, parks, transit stations, work-places, and residences. In addition to global data, Google mobility's data reports explain and detail data by continent, region, country, and city.
From the Africa-Asia report, we extracted mobility data for Senegal during 2020 and 2021, specifically for the Dakar and Thiès regions for the purposes of this brief.
%-%
\begin{table}[!h]%[!h]
\begin{center}
\caption{Data definition \citep{aurachman}\label{datadef}}
\begin{tabular}{cc}
\hline
 {\bf Classification}   & {\bf Definition}  \\
\hline
retail  & restaurant, cafes,shopping centers, parks,  \\
and recreation      & museums, libraries, movie theatres \\
\hline
grocery   & grocery markets, food warehouses, drug \\
and Pharmacies     &   stone, pharmacies, specialty, food shops,\\
 &  farmers markets\\
\hline
transit stations  & mobility trends for places like public\\
   &   transport hubs such as  subway, bus,\\
   & and train stations\\
\hline
workplaces  &  mobility trends for places of work\\
\hline
residential & mobility trends for places of residence\\
\hline
\end{tabular} 
\end{center}
\end{table}

\subsection{Analysis of Google's mobility data in Senegal}
\subsubsection{Summary statistics}
With regard to these three tables data summary (see Tables \ref{stat_20}, \ref{stat_21} and \ref{stat_22}). We observe for each year, in average and proportion, the level of representativeness of the population and the degree of their variation in an area. These values give us a lot of information. For example, we observe in 2020 that the intensity of the containment is higher because the average of "residential" is higher than the other parameters, which is not the case for the other years. In 2021, many people started to frequent the markets, magazines, shops, pharmacies, and other economic activities related to the primary needs of people because the average of "groceries and pharmacies" was higher. In addition, we could see the intensity of demand for travel by the average of "transit stations." Finally, by 2022, people had moved on to other concerns.\\
%-%
\begin{table}[!h]%[!h]
{\small
\begin{center}
\caption{Summary statistics 2020\label{stat_20}}
\begin{tabular}{ccccccccc}
\hline
 {\bf  }   & {\bf count}  &  {\bf mean} & {\bf  std}  & {\bf min} & {\bf 25\%}  & {\bf 50\%} & {\bf 75\%} & {\bf max} \\
\hline
{\bf retail and recreation}  &     1167.0	 &  -24.063410		&   14.008614    &  -64.0    	 & 	-36.0	&   -22.0    	 & 		-14.0& 28.0 \\
\hline
{\bf grocery  and pharmacies} &  1639.0   	 & -8.196461		&     15.648316  &    -61.0 	 & -19.0		&     -8.0  	 & 	1.0	&  76.0\\
\hline
{\bf parks}  &  1408.0   	 & 	-18.595881	&    15.292297   &     -62.0	 & 	-28.0	&      -16.0 	 & 	-8.0	&  30.0\\
\hline
{\bf transit stations}  &   989.0  	 & 		-32.467139  &     19.996664  &    -86.0 	 & 	-47.0	&      -31.0 	 & 	-18.0	&  10.0 \\
\hline
{\bf workplaces}  &   3436.0  	 & 	-13.671711	&  14.850636     &    -77.0  	 &       -24.0 		&     -14.0  	 & 	-2.0	& 40.0 \\
\hline
{\bf residential}  &   1228.0  	 & 	6.655537	&   6.308252    &    -15.0 	 & 		2.0  &     5.0  	 & 	10.0	&  28.0 \\
\hline
\end{tabular} 
\end{center}
}
\end{table}
%-%
\begin{table}[!h]%[!h]
{\small
\begin{center}
\caption{Summary statistics 2021\label{stat_21}}
\begin{tabular}{ccccccccc}
\hline
 {\bf  }   & {\bf count}  &  {\bf mean} & {\bf  std}  & {\bf min} & {\bf 25\%}  & {\bf 50\%} & {\bf 75\%} & {\bf max} \\
\hline
{\bf retail and recreation}  &    1231.0 	 & 	-3.211210	&   18.521424    &   -58.0  	 & 	-19.0	&    1.0   	 & 	12.0	&  54.0  \\
\hline
{\bf grocery  and pharmacies} &     1543.0	 & 	34.997408	&   25.397514    &     -40.0	 & 	15.0	&      38.0 	 & 		55.0&    152.0 \\
\hline
{\bf parks}  &     1686.0	 & 	-10.162515	&  19.455030     &    -56.0 	 & 	-19.0	&      -8.0 	 & 	1.0	&  51.0 \\
\hline
{\bf transit stations}  &  1095.0   	 & 		3.989041 &   20.994599    &     -58.0	 & 	-14.0	&    9.0   	 & 20.0		&  71.0 \\
\hline
{\bf workplaces}  &    4300.0 	 & 	-20.228372	&    16.163518   &    -85.0 	 & -31.0		&      -19.0 	 & -9.0		&  21.0\\
\hline
{\bf residential}  &    1719.0 	 & 	-2.887144	& 6.098717      &     -20.0	 & 	-6.0	&      -2.0 	 & 	1.0	& 23.0 \\
\hline
\end{tabular} 
\end{center}
}
\end{table}

%-%
\begin{table}[!h]%[!h]
{\small
\begin{center}
\caption{Summary statistics 2022\label{stat_22}}
\begin{tabular}{ccccccccc}
\hline
 {\bf  }   & {\bf count}  &  {\bf mean} & {\bf  std}  & {\bf min} & {\bf 25\%}  & {\bf 50\%} & {\bf 75\%} & {\bf max} \\
\hline
{\bf retail and recreation}  &     96.0	 & 	11.822917	&  8.798319     &  -7.0   	 & 		6.0  &    10.0   &  19.0  	 & 	 31.0 \\
\hline
{\bf grocery  and pharmacies} &     126.0	 & 	47.992063	&   15.521596    &   8.0  	 & 	39.0	&     49.5  &    58.0   	 & 	84.0  \\
\hline
{\bf parks}  &    144.0 	 & 	-6.520833	&  10.633085     &-23.0     	 & 		-15.0  &      -8.0 	 & 	0.0	& 23.0 \\
\hline
{\bf transit stations}  &   96.0  	 & 	17.375000	&   11.185752    &     -20.0	 & 	12.0	&   19.0     	 & 22.25		&  49.0  \\
\hline
{\bf workplaces}  &   370.0  	 & 	0.508108	&  17.005537     &  -48.0   	 & 	-7.0	&     2.0  	 & 	10.0	&   42.0   \\
\hline
{\bf residential}  &    150.0 	 & 	2.093333	&     4.304433  &    -10.0 	 & 0.0		&      3.0 	 & 		5.0   &   9.0  \\
\hline
\end{tabular} 
\end{center}
}
\end{table}

%-%
\begin{figure}[!h]
 	\centering
\includegraphics[width=11.5cm,height=6.cm]{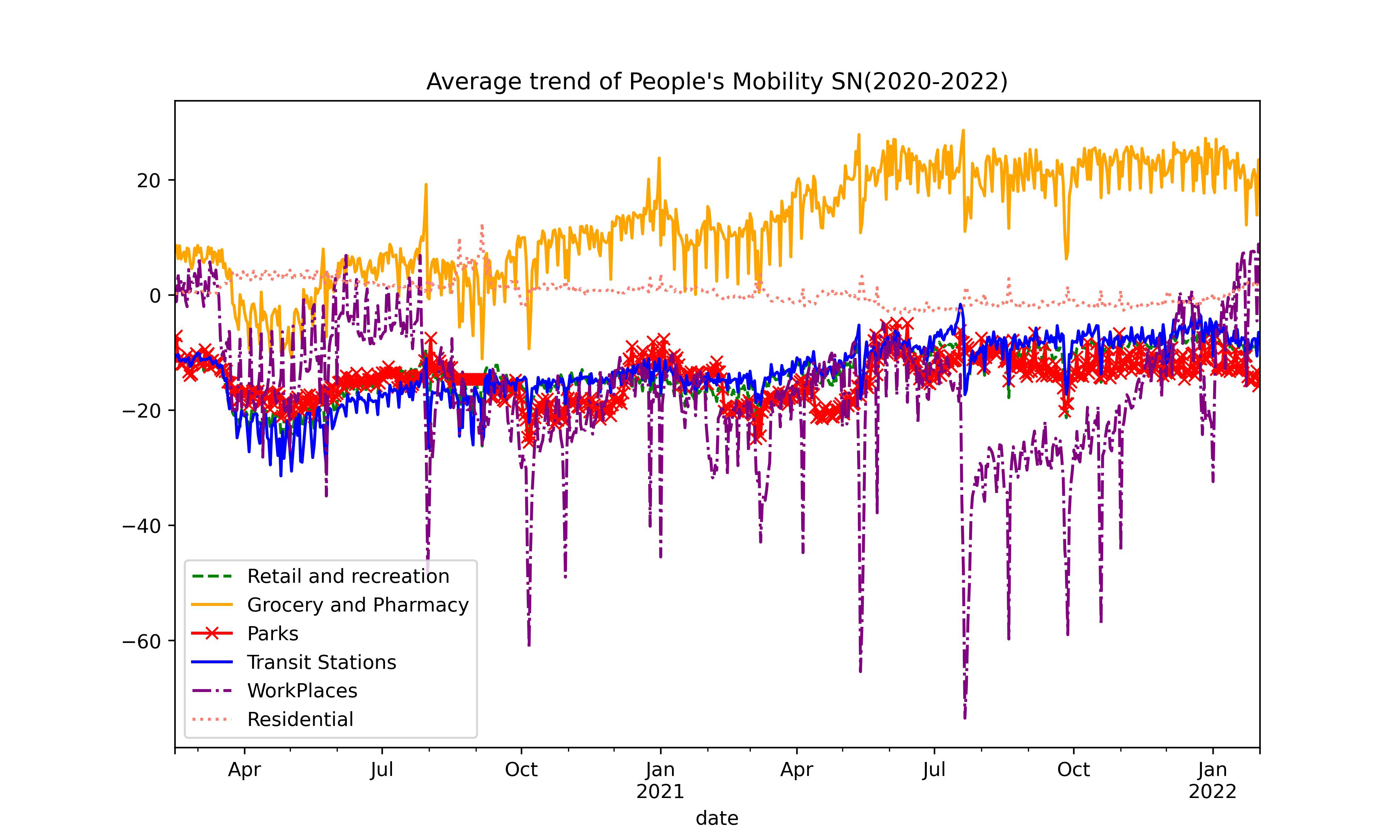}
\caption{Average mobility trend in Senegal (2020-2022)\label{average_20202022}}
 \end{figure} 	
 
\noindent The Figure \ref{average_20202022} illustrates these average trends of population mobility according to different areas; the figure visualizes in another way the information provided by summary statistics data for all three concatenated years.

%-%
\subsubsection{Average density of mobility}
Fluctuations in these densities are caused by differences in event realities and data collection.
Subject to the bias of the raw mass of information, we observe through these different figures that the density of the "Residential" zone is higher in 2020. This is mainly due to the influence of the containment, but for the rest of the years, the density of the "grocery and pharmacies" zone remains dominant. \\
This does not prevent us from carrying out our traffic prediction analysis, because during each year, people have shown up, especially since people move a lot to the activity zones regrouping food markets, food warehouses, farmers, markets, specialties, food stores, drugstores, and shopping malls. These different activities are often factors in traffic congestion when people surround themselves without any indication. Therefore, this group of activity areas is our target in the traffic forecast.
%-%
\begin{figure}[!h]
 	\centering
\includegraphics[width=5.25cm,height=5.cm]{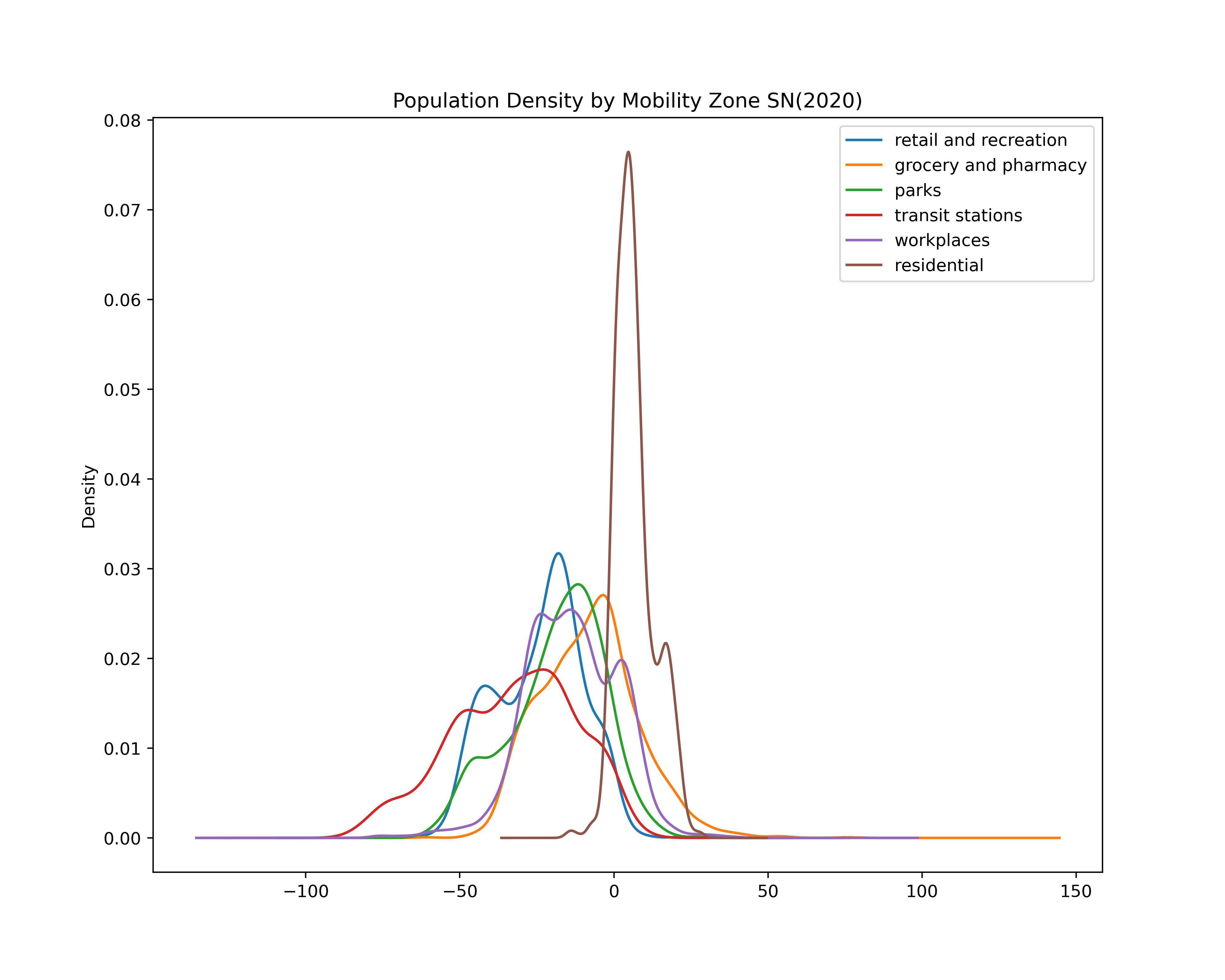}
\includegraphics[width=5.25cm,height=5.cm]{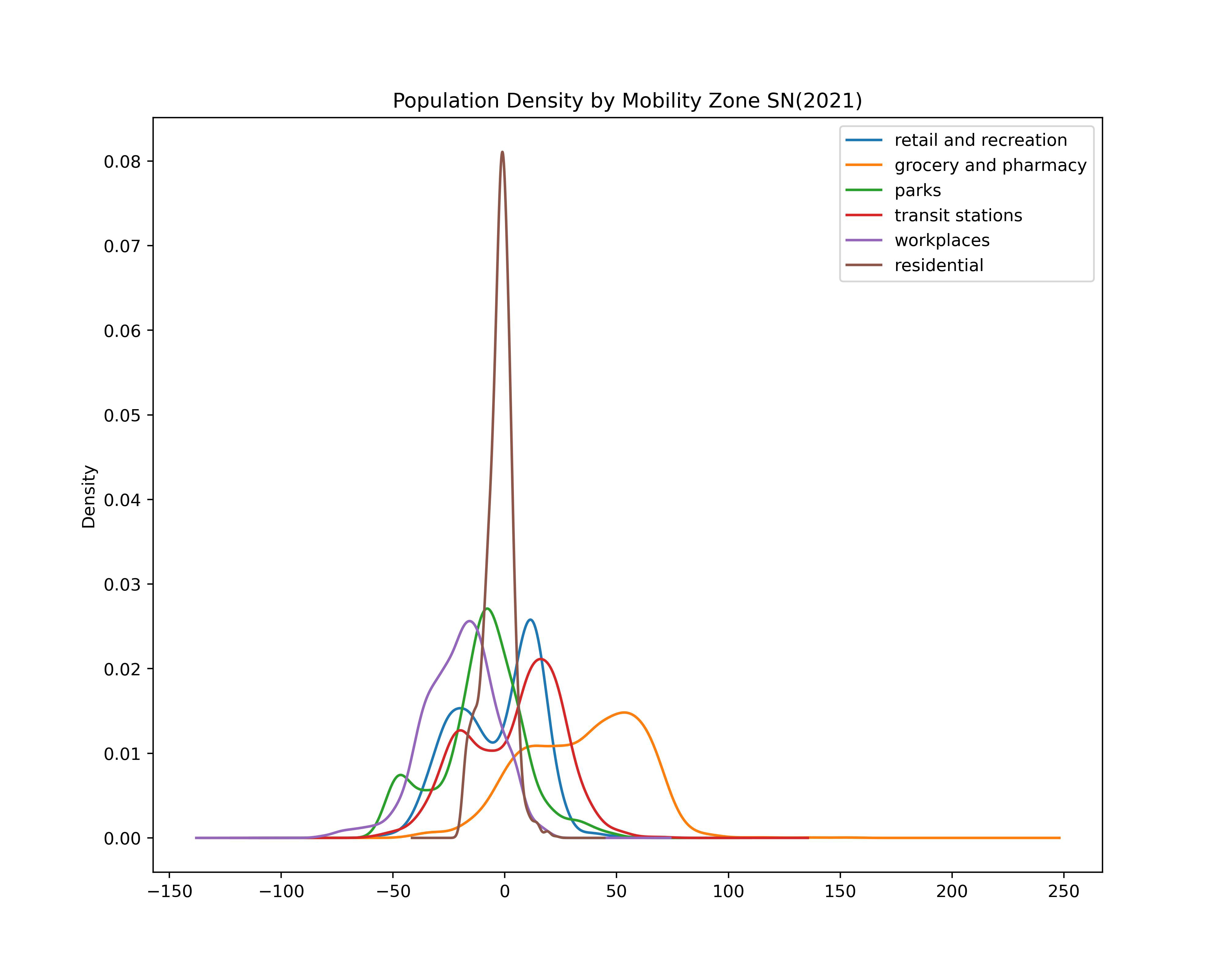}
\includegraphics[width=5.25cm,height=5.cm]{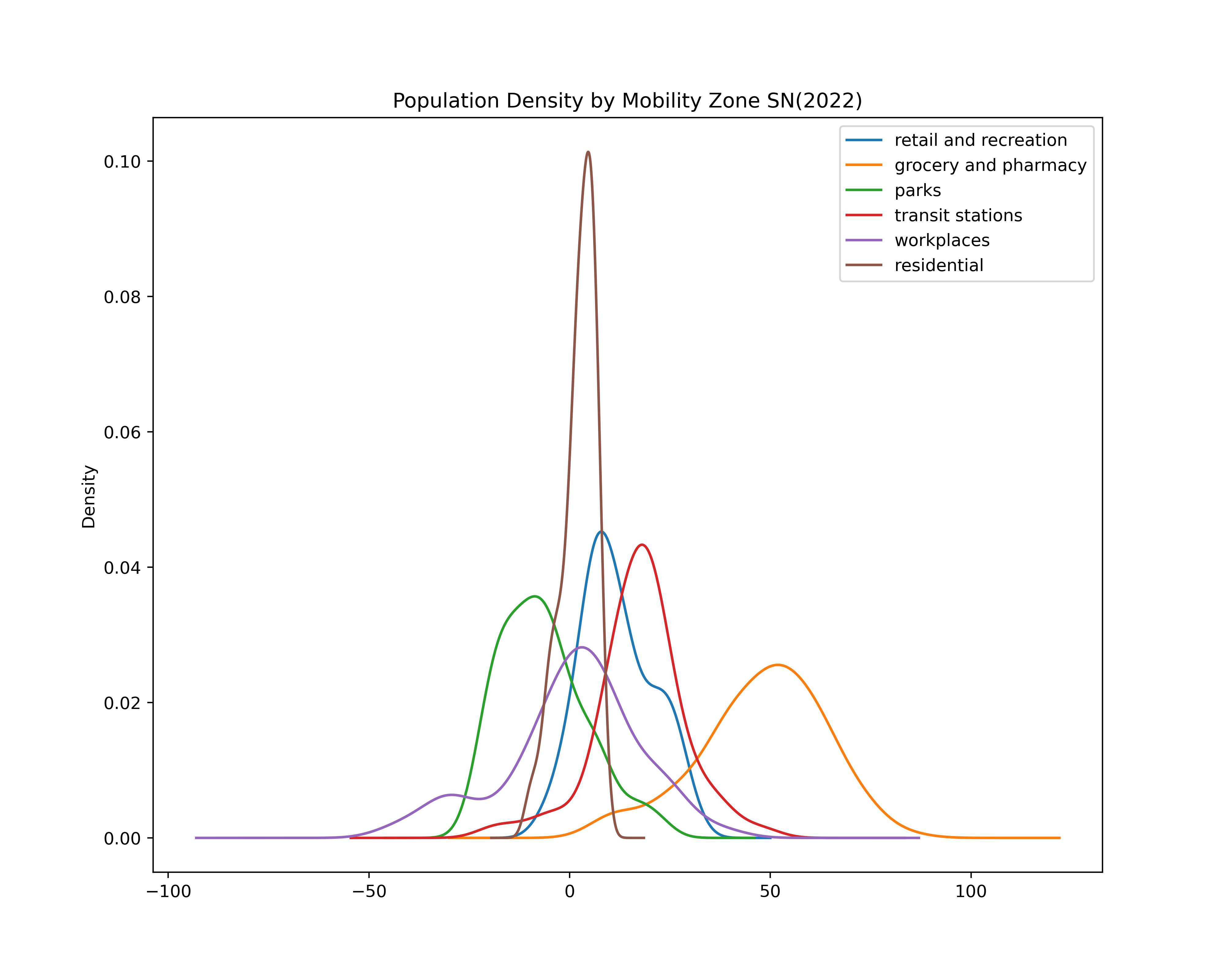}
\caption{Average density from 2020 to 2022\label{average_dens_02}}
 \end{figure}

\subsubsection{Measure of dependence}
\begin{figure}[!h]
 	\centering
\includegraphics[width=10.5cm,height=6.cm]{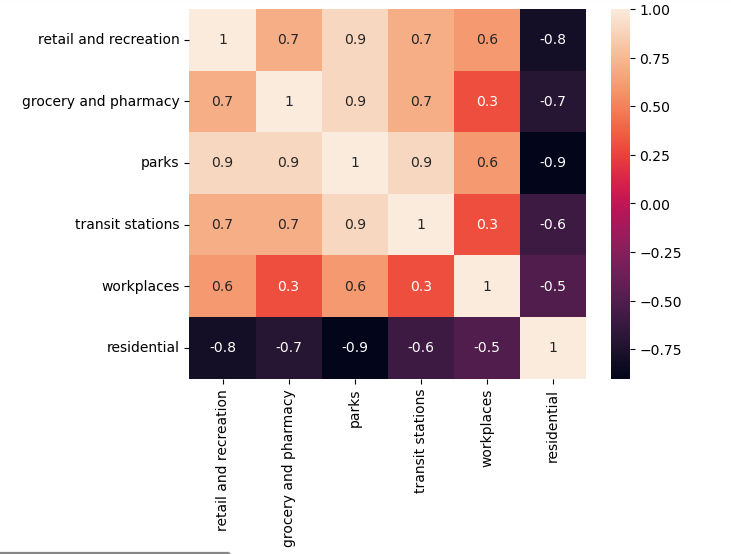}
\caption{Pairwise correlation of variables\label{Pairwise_var}}
 \end{figure}

These correlation pairs demonstrate the effects of switching from one group of activity zones to another. From this overview, we observe that the variables influence each other negatively (an increase in one leads to a decrease in the other). For example, when people leave their residences for any area, this leads to a decrease in the number of people at the residence, which explains the negative correlation of the residence variable with almost all the other variables. Only the residence variable behaves differently; the others behave in the same way. \\
Thus, this visualization help us to observe the homogeneous behavior of this group of mobility zones. In addition, this gives us the privilege of making a univariate prediction from the perspective of a factorial analysis of the data and the angle of the analysis of a time series. On this basis, we chose the "grocery and pharmacy" variable zone group for two reasons: its homogeneous behavior with the other variables and the fact that it is the densest of all the others.\\
 Its unique feature is that we can see the level of correlation for each pair of variables in proportion. Otherwise, its reading remains the same as in Figure \ref{Pairwise_var}.
%-%

\subsubsection{Regional frequency of mobility}
We observe visually, through this heat map, the extent of the phenomenon of traffic mobility
in the form of a two-dimensional color. The blue color intensity gives us visual cues on how the mobility phenomenon varies in the different regions; people during these periods were more mobile in the regions of Dakar and Thi\`es. However, in this paper, we limit the analysis to the city of Dakar.
%-%
\begin{figure}[!h]
 	\centering
\includegraphics[width=11.5cm,height=6.cm]{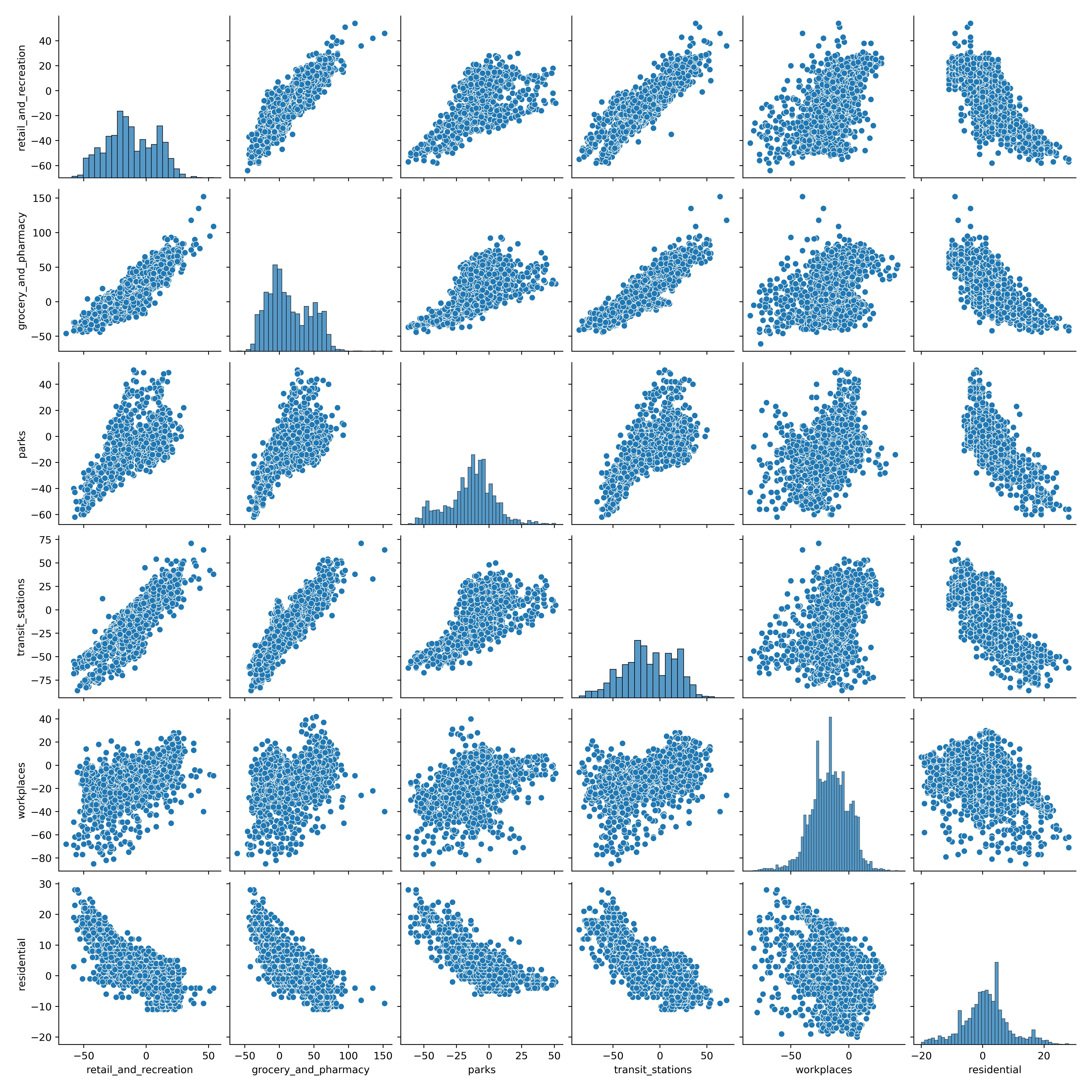}
\caption{Correlation by pair of variables\label{corr1_var}}
 \end{figure} 	
%-%
\begin{figure}[!h]
 	\centering
\includegraphics[width=11.5cm,height=5.5cm]{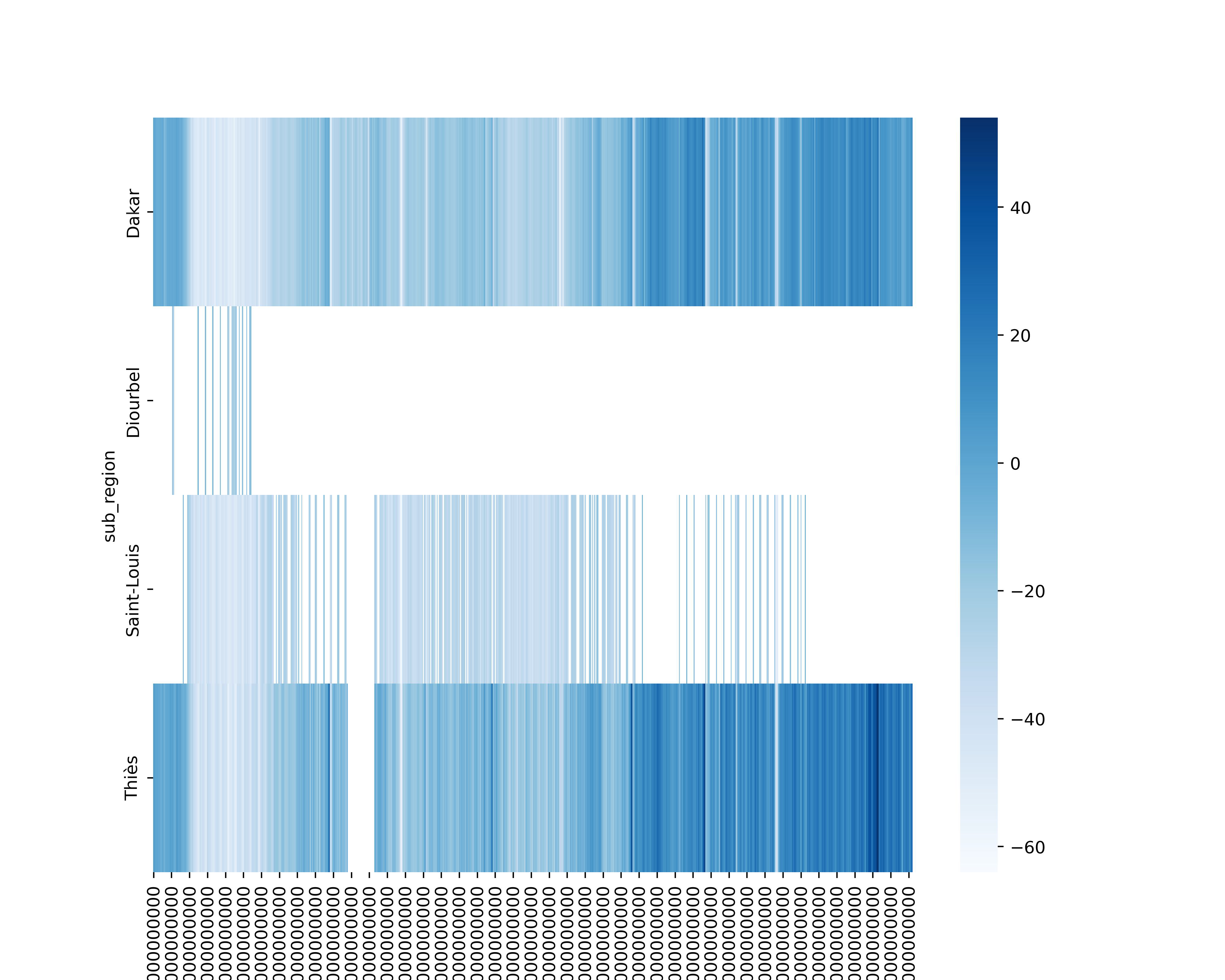}
\caption{Correlation by pair of variables\label{corr2_var}}
 \end{figure} 	

%---%
\subsection{Dakar traffic forecast using Facebook Prophet}
\subsubsection{Dakar data extraction}
As argued above, the group of grocery and pharmacy areas has a high density in all target years, which means that it also represents an important set of activities for traffic forecasting. And the Dakar region has a high mobility frequency.

%-%
\subsubsection{Analysis of the mobility time series in Dakar}
The evolution of the mobility of individuals on the variable has too many fluctuations over the years. The trend seems to be much higher in 2021 than in 2020, because the influence of the pandemic was higher in 2020 and the mobility was lower compared to 2021. In monthly and weekly averages, the trend seems to be the same, resulting in a quasi-cyclical behavior of individuals. 
This visualization projects the picture of close traffic forecast by population mobility causing the various congestion by increase in transport demand, congestion at bus stops, in traffic, etc.
%-%
\begin{figure}[!h]
 	\centering
\includegraphics[width=8cm,height=5.cm]{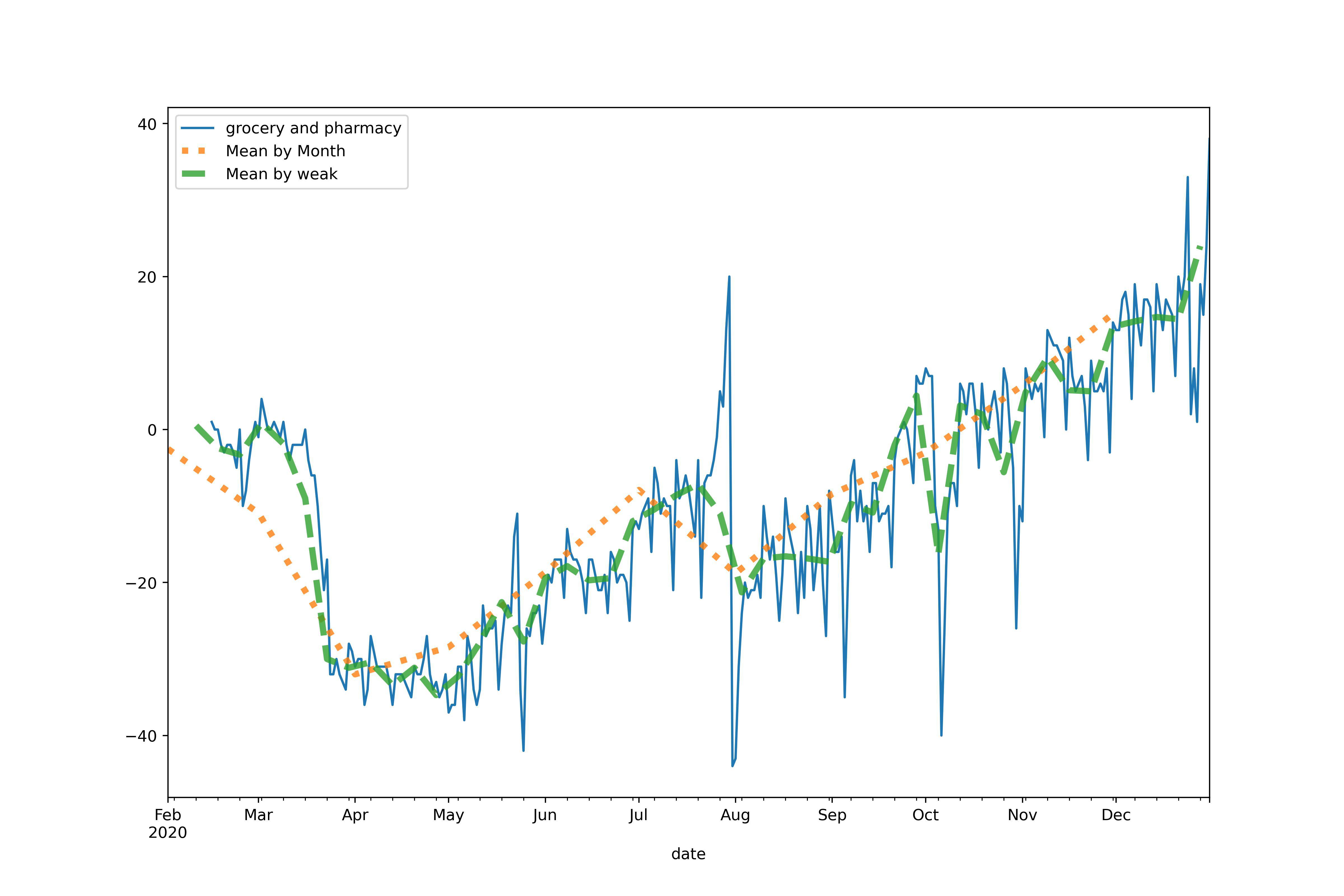}
\includegraphics[width=8cm,height=5.cm]{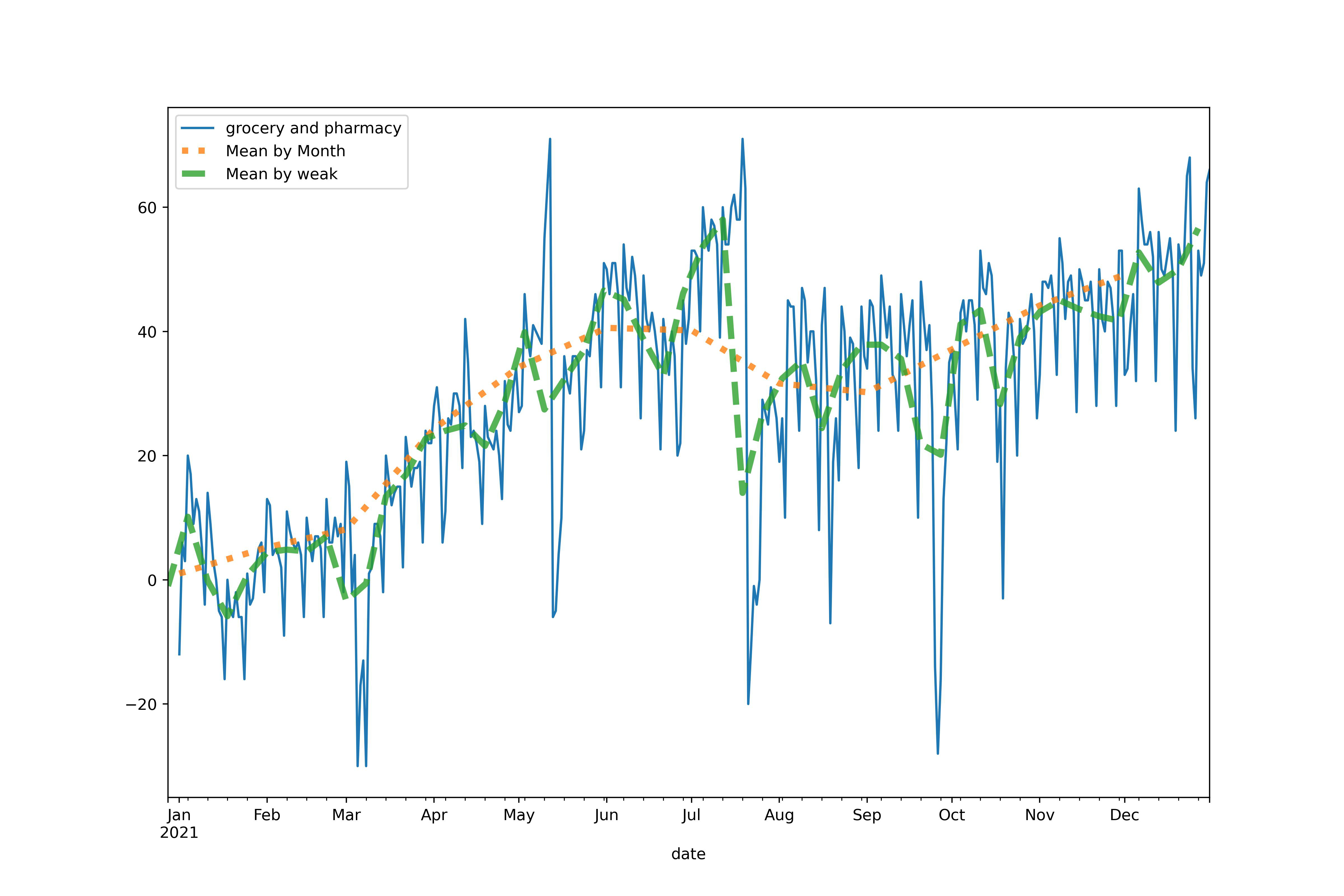}
\caption{Average evolution per weakness per month 2020 (left) and 2021 (right)\label{average_evol_w_02}}
 \end{figure}
 
\noindent In this illustration, we see the series of a weekly view that evolves with an uncertainty interval of the minimum and maximum values. The visualization serves to project the type of model to be used. The use of the additive model does not seem to be adaptable to this time series because the lines of the uncertainty interval are not parallel.
%-%
\begin{figure}[!h]
 	\centering
\includegraphics[width=10.5cm,height=6.cm]{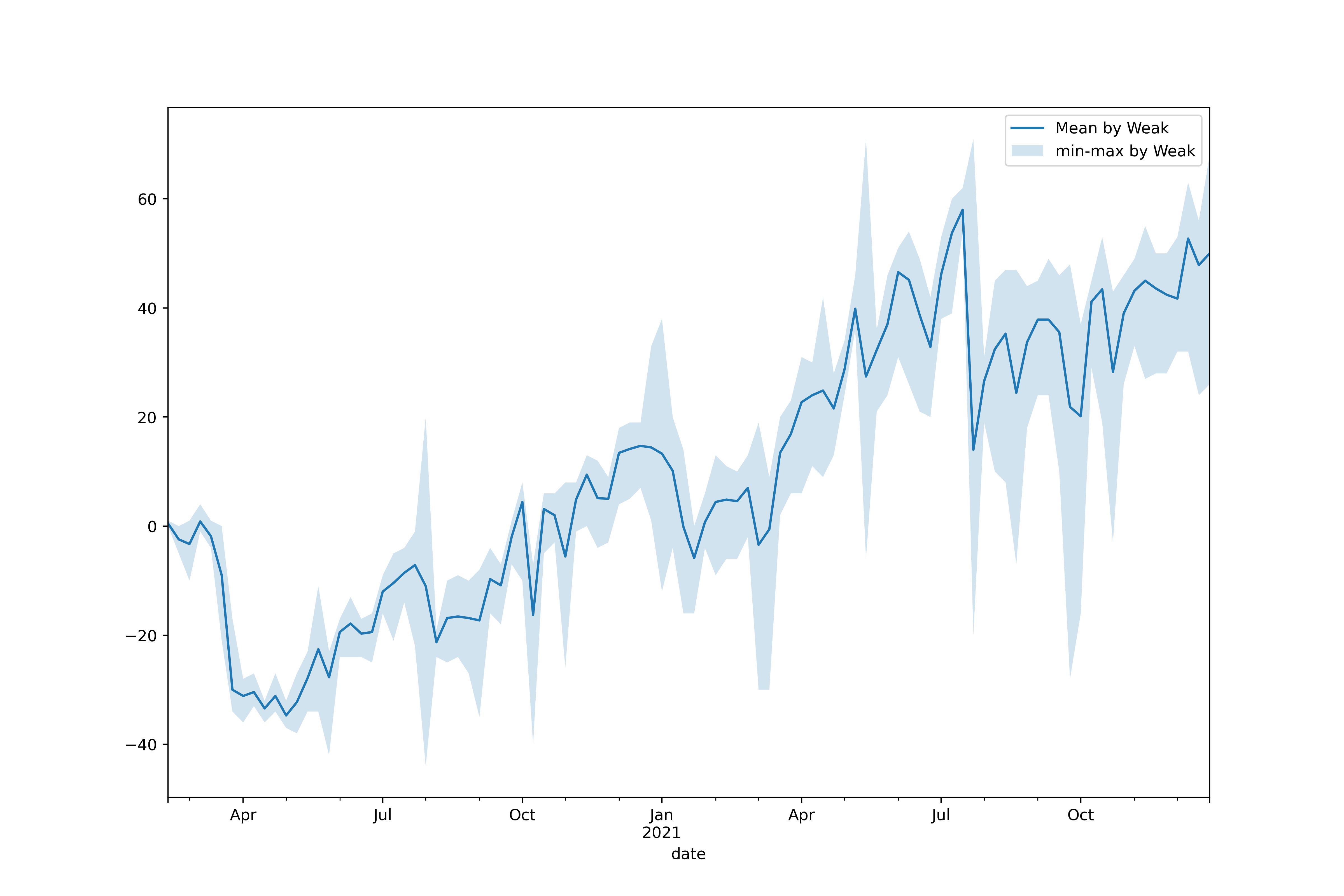}
\caption{Weekly trend with a zone of uncertainty\label{weekly_t}}
 \end{figure} 	

\subsubsection{Data preparation}
\begin{itemize}
\item {\bf Data extraction}: Apart from extracting data from Africa, Senegal, and Dakar in the initial dataset drawn from the Google Africa-Asia mobility data detail report source \ref{googdata}. We highlighted the grocery and pharmacy action area groups to use as a target for our traffic prediction study.
\item  {\bf Data missing}: Our sample dataset has no missing values, although Prophet is well resilient to missing data, i.e., the model works even if there is missing data, but this does not cause the model to be insensitive when learning, so there is a probability that the prediction may be biased. Therefore, Facebook experts suggest deletion as
an optimal solution. 
\item  {\bf Column naming}: We renamed the date column to ds (the timestamp column) and the grocery and pharmacy column to y (the target column), because Prophet's input is always a frame of data: $ds$ and $y$.
\end{itemize}
%-%
\begin{table}[!h]%[!h]
{\small
\begin{center}
\caption{Dakar data preparation for training\label{data_dk_prep}}
\begin{tabular}{ccc}
\hline
 {\bf  }   & {\bf ds}  &  {\bf y}\\
\hline
{\bf 0}  &  2020-02-15  	 & 	1.0	 \\
\hline
{\bf 1}  &  2020-02-16    	 &   0.0		 \\
\hline
{\bf 2}  &   2020-02-17   	 & 	0.0	 \\
\hline
{\bf 3}  &   2020-02-18   	 & 	-2.0	 \\
\hline
{\bf 4}  &   2020-02-19   	 &   -3.0 \\
\hline
{\bf ...}  &     ...	 & ...	\\
\hline
{\bf 681}  &   2021-12-27   	 & 53.0	\\
\hline
{\bf 682}  &   2021-12-28  	 &  49.0	\\
\hline
{\bf 683}  &   2021-12-29  	 & 51.0	\\
\hline
{\bf 684}  &   2021-12-30  	 & 64.0 	\\
\hline
{\bf 685}  &   2021-12-31  	 & 66.0	\\
\hline
\end{tabular} 
\end{center}
}
\end{table}

\subsubsection{Simple model prediction}
\noindent {\bf Data training.}\\ 
The model is adapted by instantiating a new object called "Prophet." For this model, we use the default hyperparameters and train the model by the fit method. We started by training a simple Prophet model, i.e., using the default hyperparameters. And we re-trained this model with the definition of some custom hyperparameters in order to improve the model.  \\

\noindent {\bf Data forecasting.}\\
The Prophet model provides a new dataset to be predicted by the "make future dataframe" method while defining a quarterly forecast period with daily frequency.  Then, we predict by using the "predict" method which assigns to each future line a predicted value.  This is stored in an object that forms a new database containing also the information for all components of the model and the uncertainty intervals. and the uncertainty intervals. \\
\noindent Let: mt = multiplicative\_terms and ad = additive\_terms.\\
%-%
\begin{table}[!h]%[!h]
{\tiny
\begin{center}
\caption{Data forcasting simple model(1)\label{forcasting_1}}
\begin{tabular}{ccccccccccc}
\hline
 {\bf ds  }   & {\bf trend}  &  {\bf yhat\_lower}  & {\bf yhat\_upper}    &  {\bf trend\_lower}     & {\bf trend\_upper}    & {\bf daily}   & {\bf daily\_lower}     & {\bf daily\_upper}  &  {\bf holidays}  &  {\bf ...} \\
\hline
        2022-03-27       &     69.053300   &     22.305144    &   43.536848    &  64.955904 &  72.961736      &  -0.102652    &  -0.102652    &    -0.102652    &    0.0    &   ...    \\
\hline
        2022-03-28       &     69.121184   &    47.667193    &   68.617346    &  64.965781 &  73.083959      &  -0.102652    &  -0.102652    &    -0.102652    &    0.0    &   ...    \\
\hline
        2022-03-29       &     69.189068   &     42.387100    &   64.503059    &   64.975658    &   73.207121 &    -0.102652    &  -0.102652    &    -0.102652    &    0.0    &   ...    \\
\hline
        2022-03-30       &    69.256951    &  37.454944     &   58.414284   &  64.985535    &   73.383817 &    -0.102652    &  -0.102652    &    -0.102652    &    0.0    &   ...    \\
\hline
        2022-03-31       &    69.324835    &  59.112695     &  82.227243   &  64.990824    &   73.528296 &    -0.102652    &  -0.102652    &    -0.102652    &    0.0    &   ...    \\

\hline
\end{tabular} 
\end{center}
}
\end{table}
\par
\begin{figure}[!h]
 	\centering
\includegraphics[width=8cm,height=5.cm]{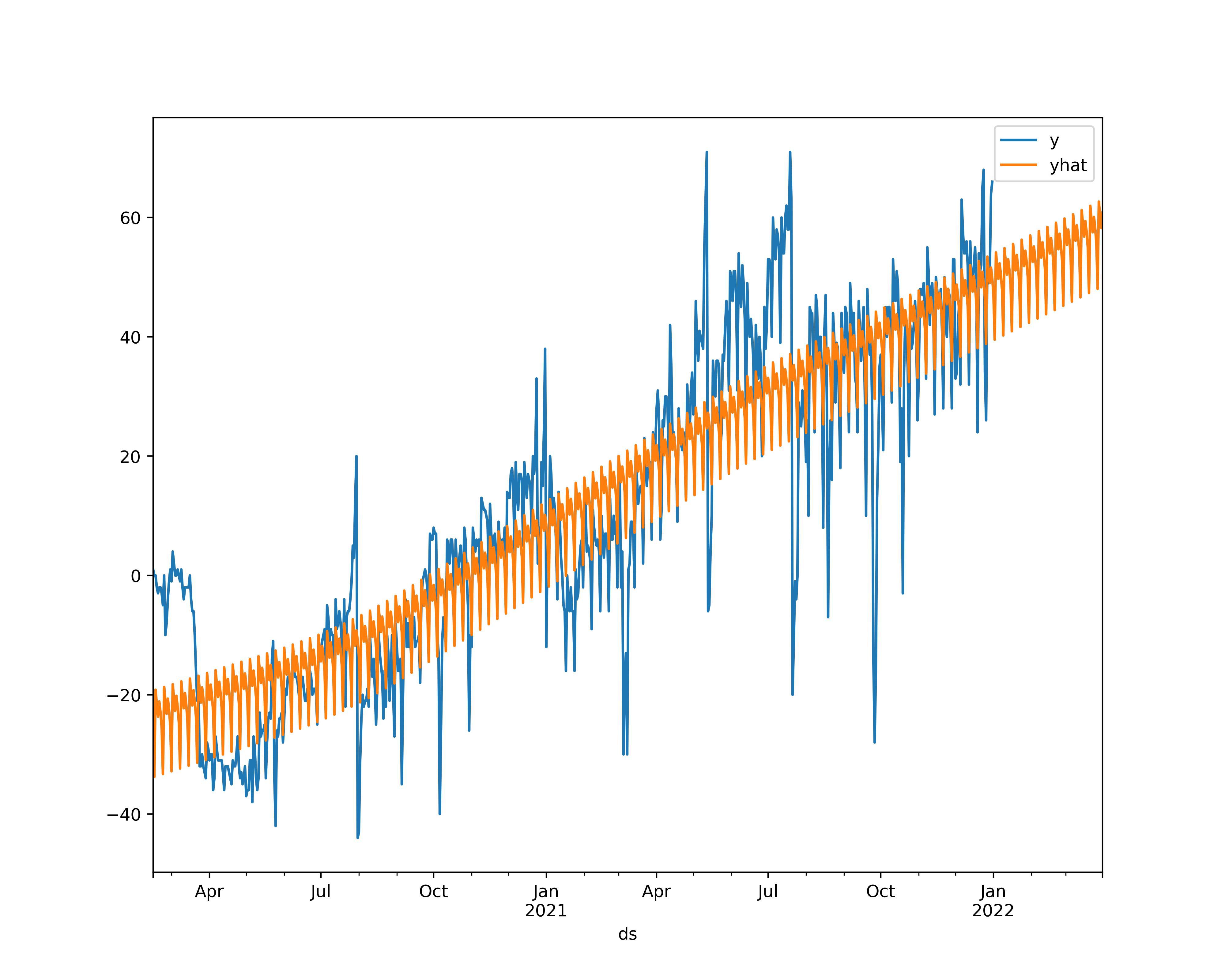}
\includegraphics[width=8cm,height=5.cm]{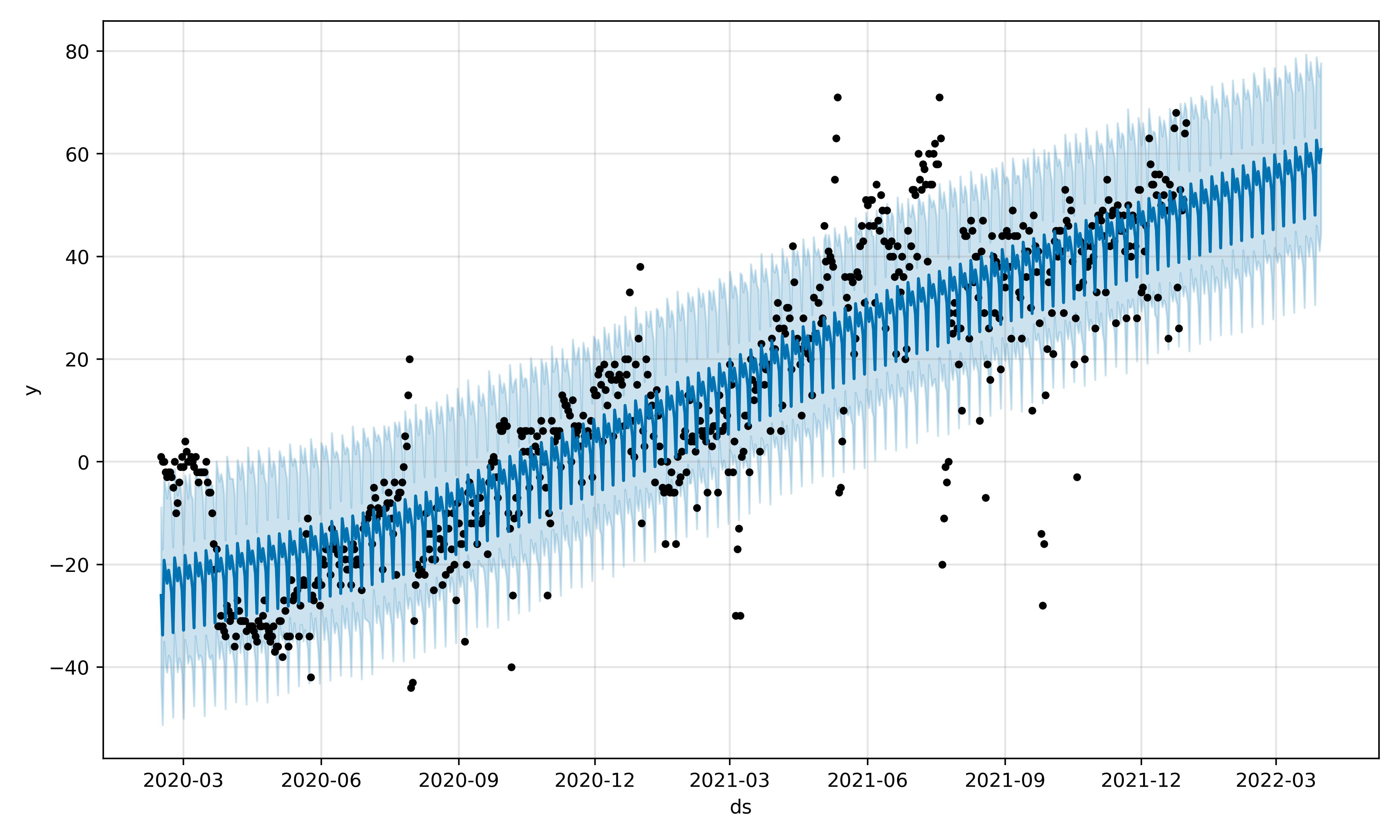}
\caption{Simple model prediction\label{Sim_m_pred}}
 \end{figure}

\vspace{-1.cm}
%-%
\begin{table}[!h]%[!h]
{\scriptsize
\begin{center}
\caption{Data forcasting simple model(2)\label{forcasting_2}}
\begin{tabular}{ccccccccc}
\hline
 {\bf mt\_lower  }   & {\bf mt\_upper}  &  {\bf weekly}  & {\bf weekly\_lower}    &  {\bf weekly\_upper}     & {\bf ad}    & {\bf ad\_lower}  & {\bf ad\_upper}  & {\bf yhat}\\
\hline
      -0.519573        &   -0.519573     &   -0.242731    &    -0.242731    &  -0.242731      &     0.0   &   0.0 &  0.0    &     33.175070    \\
\hline
      -0.162073        &  -0.162073      &      0.141448   &   0.141448    &    0.141448    &    0.0    &  0.0   &  0.0  &    57.918514     \\
\hline
-0.226271         &   -0.226271      &     0.042683    &  0.042683     & 0.042683       &     0.0   &    0.0   &  0.0  &   53.533577      \\
\hline
       -0.304403       &     -0.304403    &    -0.120948      &  -0.120948     &   -0.120948     &     0.0   &   0.0      & 0.0     &     48.174911    \\
\hline
        0.013417      &   0.013417      &    0.100513     &    0.100513   &   0.100513     &      0.0  &  0.0     &  0.0  &      70.254973   \\
\hline
\end{tabular} 
\end{center}
}
\end{table}
\noindent {\bf The prediction of the simple model.}\\
The predicted model is illustrated in two different figures (see Figure \ref{Sim_m_pred}). The blue curve in the first figure represents the initial series, while the orange curve represents the predictive series.\\
In the second figure, the points represent the training base, and the light blue area surrounding the predicted series in dark blue is the uncertainty layer. This expresses the confidence interval that can be estimated around our prediction. \\
As Prophet uses an additive model by default, we can visibly observe the intensity of the prediction errors. The model does not really adjust for seasonality. This is why we will adjust or improve it. \\
%-%

\noindent {\bf Performance of the simple model.}\\
We measure the performance by observing the prediction error through the cross-validation feature. We specify the prediction horizon (horizon), the size of the initial training period (initial), and the spacing between the cut-off dates (period). By default, the initial training period is set to three, but here we have 680 days with an estimation horizon of 5 days.\\
The cross-validation produces a database containing the actual $y$-values and off-sample forecast values $yhat$ for each simulated forecast date and cut-off date. In particular, a forecast is made for each observed point between the cutoff and the horizon. This data frame depicts an estimate made by the model of the horizon last 5 days of our Data Initial, which is the date of December 27 to December 31, 2021.
%-%
\begin{table}[!h]%[!h]
{\small
\begin{center}
\caption{Simple cross-validation model\label{s_cross_val}}
\begin{tabular}{ccccccc}
\hline
            & {\bf ds}  &  {\bf yhat} & {\bf yhat\_lower }  & {\bf yhat\_power} & {\bf y} & {\bf cutoff} \\
\hline
{\bf 0}  &     2021-12-27	 & 	53.628160	&  35.329445     &  70.867974    	 & 53.0		&      2021-12-26  \\
\hline
{\bf 1}  &     2021-12-28	 & 	51.328594	&  35.837800     &    67.427711	  & 		49.0    &      2021-12-26   \\
\hline
{\bf 2}  &     2021-12-29	 & 	49.183583	&     32.151436  &    65.300858 	 & 	51.0	&   2021-12-26      \\
\hline
{\bf 3}  &   2021-12-30  	 & 	51.666740	&    35.440260   &    69.541234 	 & 	64.0   &   2021-12-26      \\
\hline
{\bf 4}  &  2021-12-31   	 & 	49.809984	&  33.272027     &    69.625405	 & 	66.0   &   2021-12-26      \\

\hline
\end{tabular} 
\end{center}
}
\end{table}
\noindent As an example, we can look at the largest deviation from prediction here, which occurs on the last day, with the true value $y$ = 66.0 and the predicted value $yhat$ = 49.809984. The error is about 24.5 percent, according to the performance measures mape and mdape (see Table \ref{s_cross_val}).
%-%
\begin{table}[!h]%[!h]
{\small
\begin{center}
\caption{Simple model performance measurement\label{meas_perf}}
\begin{tabular}{ccccccccc}
\hline
 {\bf  }   & {\bf horizon}  &  {\bf mse} & {\bf  rmse}  & {\bf mae} & {\bf mape}  & {\bf mdape} & {\bf smape} & {\bf coverage} \\
\hline
{\bf 0}  &    1 days	 & 	0.394585	&    0.628160     &    0.628160 	 & 	0.011852	&   0.011852     & 	0.011782	&    1.0  \\
\hline
{\bf 1}  &    2 days 	 & 	5.422351	&    2.328594     &    2.328594 	 & 	0.047522	&    0.047522     & 	0.046419	&   1.0   \\
\hline
{\bf 2}  &   3 days   	 & 	3.299369	&     1.816417   &     1.816417	 & 	 0.035616 &   0.035616     &      0.036262     &   1.0     \\
\hline
{\bf 3}  &   4 days  	 & 	152.109312    &   12.333260    &     12.333260   &  0.192707	&    0.192707     & 	0.213255	&    1.0    \\
\hline
{\bf 4}  &  5 days    	 & 	262.116604	&    16.190016   &   16.190016  	 & 	0.245303	&   0.245303     & 	0.279596	&   1.0        \\

\hline
\end{tabular} 
\end{center}
}
\end{table}
\noindent All performance measures of this simple model have an increasing trend, which means that the error is increasing and the model is not performing well. 
All these measures were calculated using a "performance metrics" feature used for prediction performance ($yhat$, $yhat\_lower$, and $yhat\_upper$ vs. $y$) as a function of distance from threshold. 
The statistics calculated are the root mean square error (RMSE), mean absolute error (MAE), mean absolute percentage error (MAPE), median absolute percentage error (MDAPE), and coverage of the estimates "$yhat\_lower$". \\
The values "$yhat\_lower$" and "$yhat\_upper$" are calculated over a sliding window of the predictions in our dataset provided by cross-validation.
%-%
\begin{figure}[!h]
 \centering
\includegraphics[width=11.5cm,height=6.5cm]{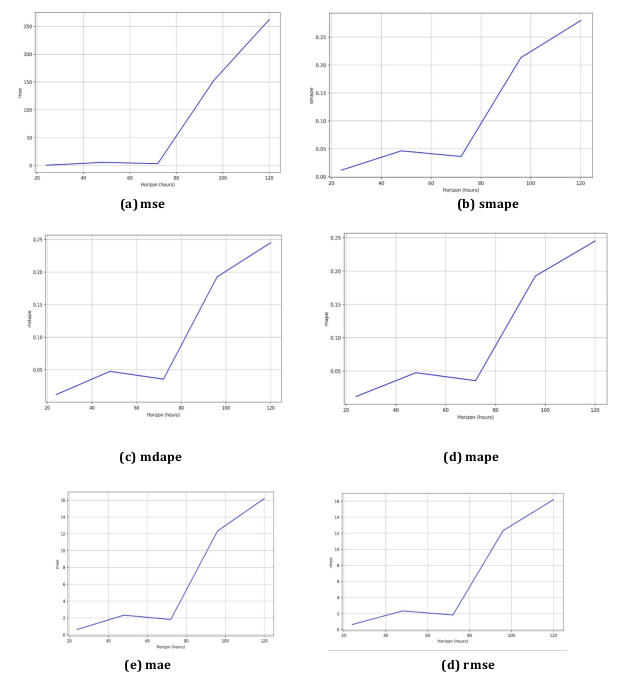}
\caption{Simple model performance measures\label{simpl_m_pe}}
 \end{figure} 	

\subsubsection{Improved model prediction}
\noindent {\bf Definition of the adjusted hyperparameters.}\\
Some elements helped in the adjustment of the model, such as:
\begin{itemize}
\item Trend change points: allows the trend to be adjusted appropriately; if we want to customize it, we can use the n\_changepoint or changepoint\_range arguments. And to keep To balance the flexibility of the trends, i.e., avoid over-fitting (too much flexibility) and under-fitting (not enough flexibility), we set the argument "change-point\_prior\_scale" to 0.5 (the default value).
\item The multiplicative model is used. We are confronted with multiplicative seasons that evolve in lockstep with the trend. Despite its default additive modeling, Prophet can
model multiplicative seasonality by passing the argument "seasonality\_mode=multiplicative." It has also been used to adjust for seasonality. The simple model based on the additive model is not better suited to the type of data that has multiple seasonality.
\item Uncertainty intervals in the trend are defined at 0.95 using the argument "interval\_width," by full Bayesian sampling using the argument "mcmc.sample," and are defined at 300.
\item Vacation customization: Since our series also covers a pendulum period in its learning base, we will need to avoid having large dips and peaks captured by the trend component. On this, we treated the days affected by covid-19 as holidays that will not be repeated in the future.
\end{itemize}
We have an entry for each containment period, with ds specifying the start of containment, ds\_upper is not used by Prophet, but it is a convenient way for us to compute upper window. And upper\_window tells Prophet that the containment extends for $x$ days after the containment starts. Note that the holiday regression includes the upper bound, see Tables \ref{V_per}, \ref{d_for_imp1} and \ref{d_for_imp2}.
\par\vspace{-.25cm}
%-%
\begin{table}[!h]%[!h]
{\small
\begin{center}
\caption{Vacation period\label{V_per}}
\begin{tabular}{cccccc}
\hline
 {\bf  }   & {\bf horizon}  &  {\bf ds} & {\bf  lower\_window}  & {\bf ds\_upper} & {\bf upper\_window} \\
\hline
{\bf 0}  &    lockdown\_1  & 	2020-03-21	&    0    &   2020-06-06  	 & 	  77   \\
\hline
{\bf 1}  &    lockdown\_2 	 & 	2020-07-09	&    0   &   2020-10-27  	 & 110		 \\
\hline
{\bf 2}  &   lockdown\_3  	 &  2021-02-13 		&    0   &   2021-02-17  	 & 	4     \\
\hline
{\bf 3}  &   lockdown\_4	 &   2021-05-28		&   0    &    2021-06-10 	 & 	13	        \\
\hline
\end{tabular} 
\end{center}
}
\end{table}
\par\vspace{-1.0cm}
%-%
\begin{table}[!h]%[!h]
{\tiny
\begin{center}
\caption{Data forecasting improved model(1)\label{d_for_imp1}}
\begin{tabular}{ccccccccccc}
\hline
 {\bf ds }   & {\bf trend}  &  {\bf yhat\_lower}  & {\bf yhat\_upper}    &  {\bf trend\_lower}     & {\bf trend\_upper}    & {\bf daily}  & {\bf daily\_lower}  & {\bf daily\_upper}   & {\bf holidays}    & {\bf ...} \\
\hline
    2022-03-27     &     69.053300   &  23.324276   &   43.671500    &  64.835821      &  73.351689      &  -0.102652    &  -0.102652    &  -0.102652    & 0.0     &  ...   \\
    \hline
    2022-03-28     &    69.121184    &  45.898250    &  68.572174    &    64.848931    &   73.473789    &  -0.102652    &  -0.102652    &   -0.102652   &  0.0    &   ...  \\
\hline
    2022-03-29     &    69.189068    &   42.293578     &  64.307162     &   64.866176     &  73.595889      &  -0.102652    &  -0.102652    &  -0.102652    &   0.0   &  ...   \\
\hline
    2022-03-30     &    69.256951    &    37.270085     &  58.927286     &  64.881536      &   73.721796     &   -0.102652   &  -0.102652    &  -0.102652    &   0.0   &  ...   \\
\hline
    2022-03-31     &    69.324835    &    59.672033     &  81.474296     &  64.893011   &  73.858335     &   -0.102652   &  -0.102652    &  -0.102652    &   0.0   &   ...  \\
\hline
\end{tabular} 
\end{center}
}
\end{table}
\par
\begin{figure}[!h]
 	\centering
\includegraphics[width=8cm,height=5.cm]{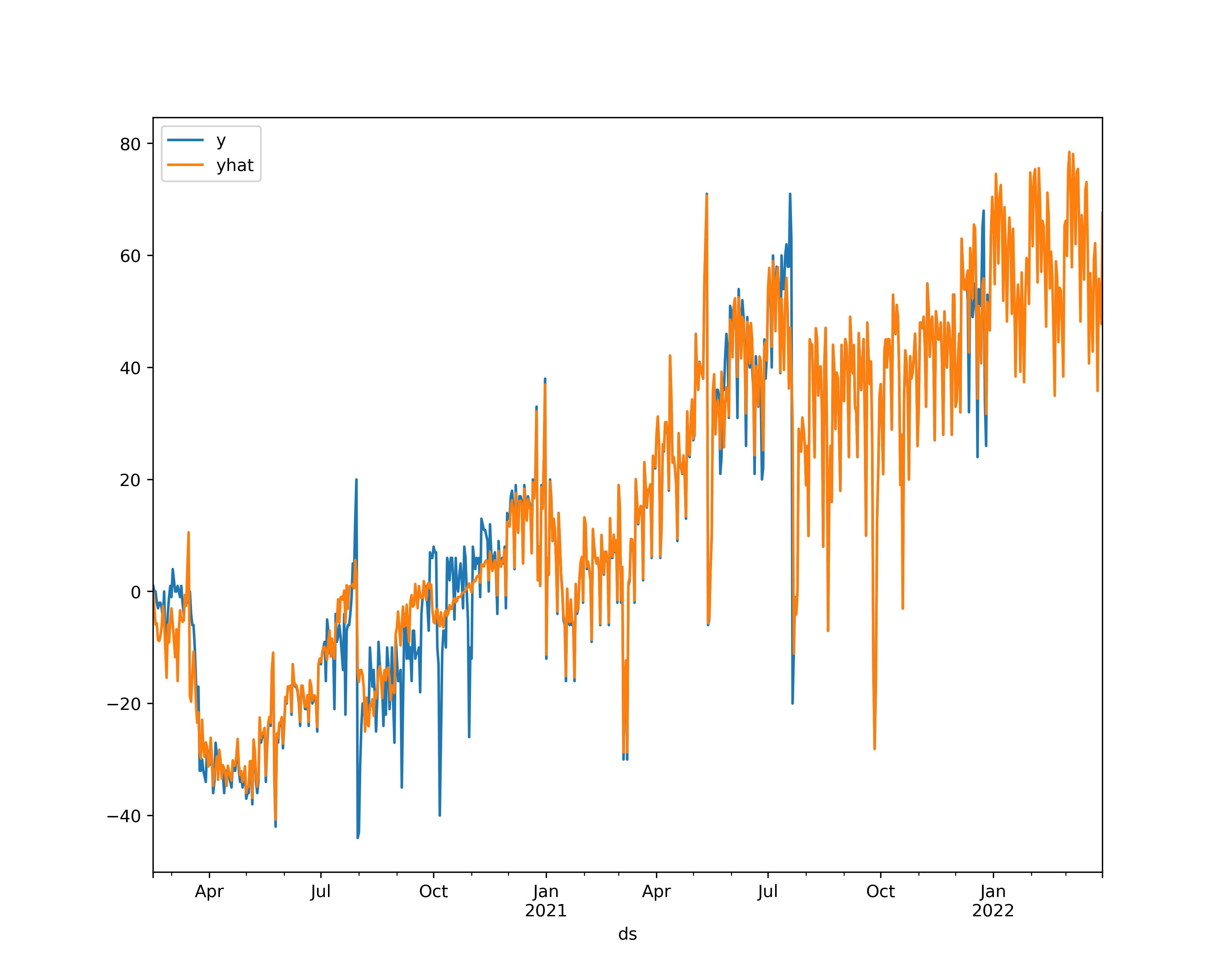}
\includegraphics[width=8cm,height=5.cm]{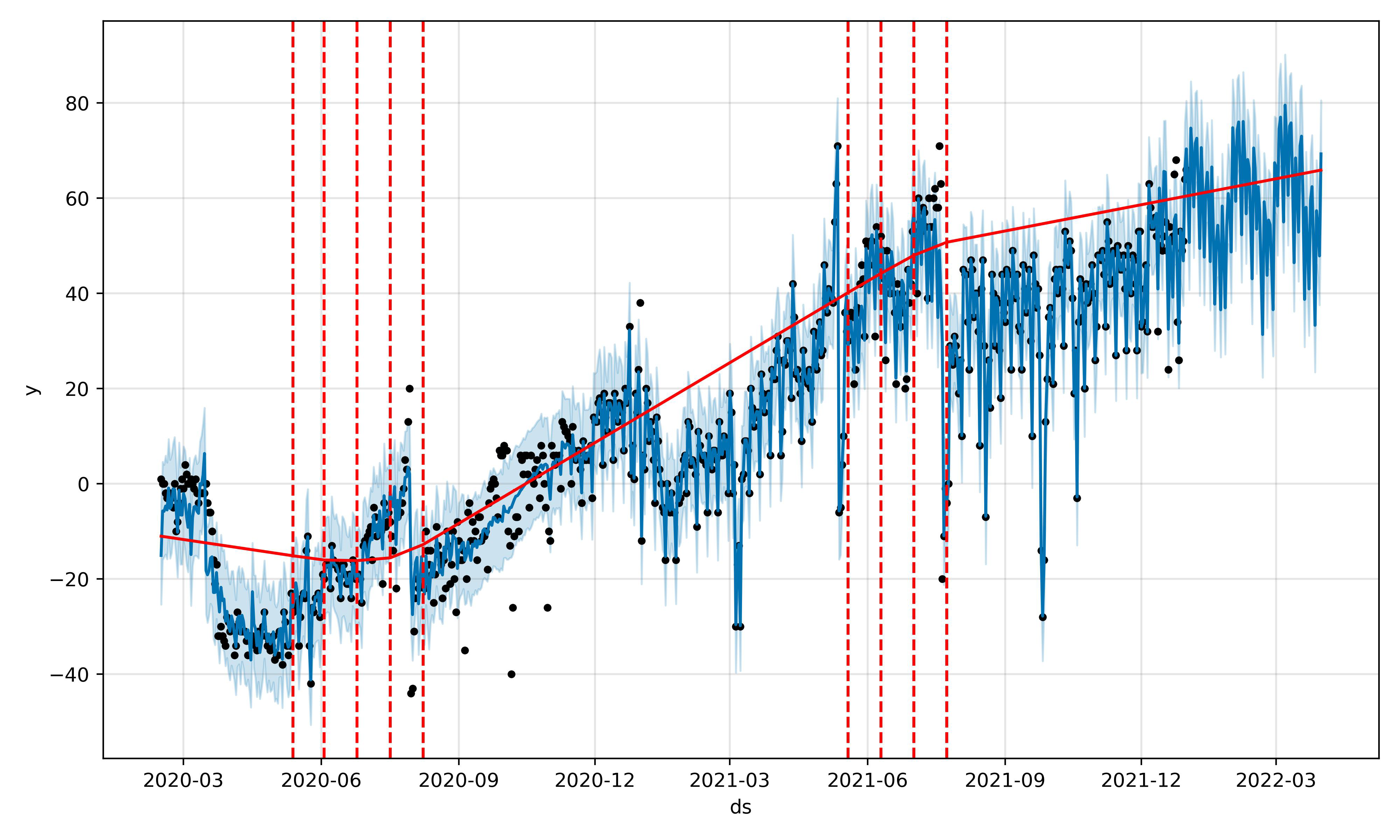}
\caption{Simple model prediction\label{Simp_mo_pred}}
 \end{figure}

\vspace{-1.0cm}
%-%
\begin{table}[!h]%[!h]
{\scriptsize
\begin{center}
\caption{Data forecasting improved model(2)\label{d_for_imp2}}
\begin{tabular}{ccccccccc}
\hline
 {\bf mt\_lower }   & {\bf mt\_upper}  &  {\bf weekly}  & {\bf weekly\_lower}    &  {\bf weekly\_upper}     & {\bf ad}    & {\bf ad\_lower } & {\bf ad\_upper}  & {\bf yhat}   \\
    \hline
 -0.519573      &   -0.519573      &      -0.242731   &  -0.242731    &   -0.242731     &     0.0   &    0.0   &   0.0   &    33.175070  \\
\hline
      -0.162073    &   -0.162073     &   0.141448       &  0.141448     &   0.141448     &     0.0   &    0.0   &   0.0   &   57.918514   \\
\hline
   -0.226271      &     -0.226271      &  0.042683      &  0.042683     &   0.042683      &     0.0   &    0.0   &   0.0   &   53.533577   \\
\hline
    -0.304403     &      -0.304403      &      -0.120948       &   -0.120948    &   -0.120948     &     0.0   &    0.0   &   0.0   &   48.174911   \\
\hline
    0.013417     &      0.013417      &      0.100513       &   0.100513    &   0.100513     &     0.0   &    0.0   &   0.0   &   70.254973   \\
\hline
\end{tabular} 
\end{center}
}
\end{table}
%\ \\
\noindent {\bf The prediction of the improved model.}\\
The improved model prediction is also illustrated in two different  figures. The blue curve in Figure \ref{Simp_mo_pred} illustrates the initial series, and the orange one shows the predicted series.  
In the second figure, the points represent the training base. The dark  blue curve represents the predicted series; the light blue layer that  surrounds it represents the uncertainty zone, which expresses the  confidence interval that we can estimate around our prediction; and 
the red line represents the trend prediction. The red dotted lines   represent the cuts at the trend changing points.
%-%
%-%
\begin{figure}[!h]
 	\centering
\includegraphics[width=9.5cm,height=8.5cm]{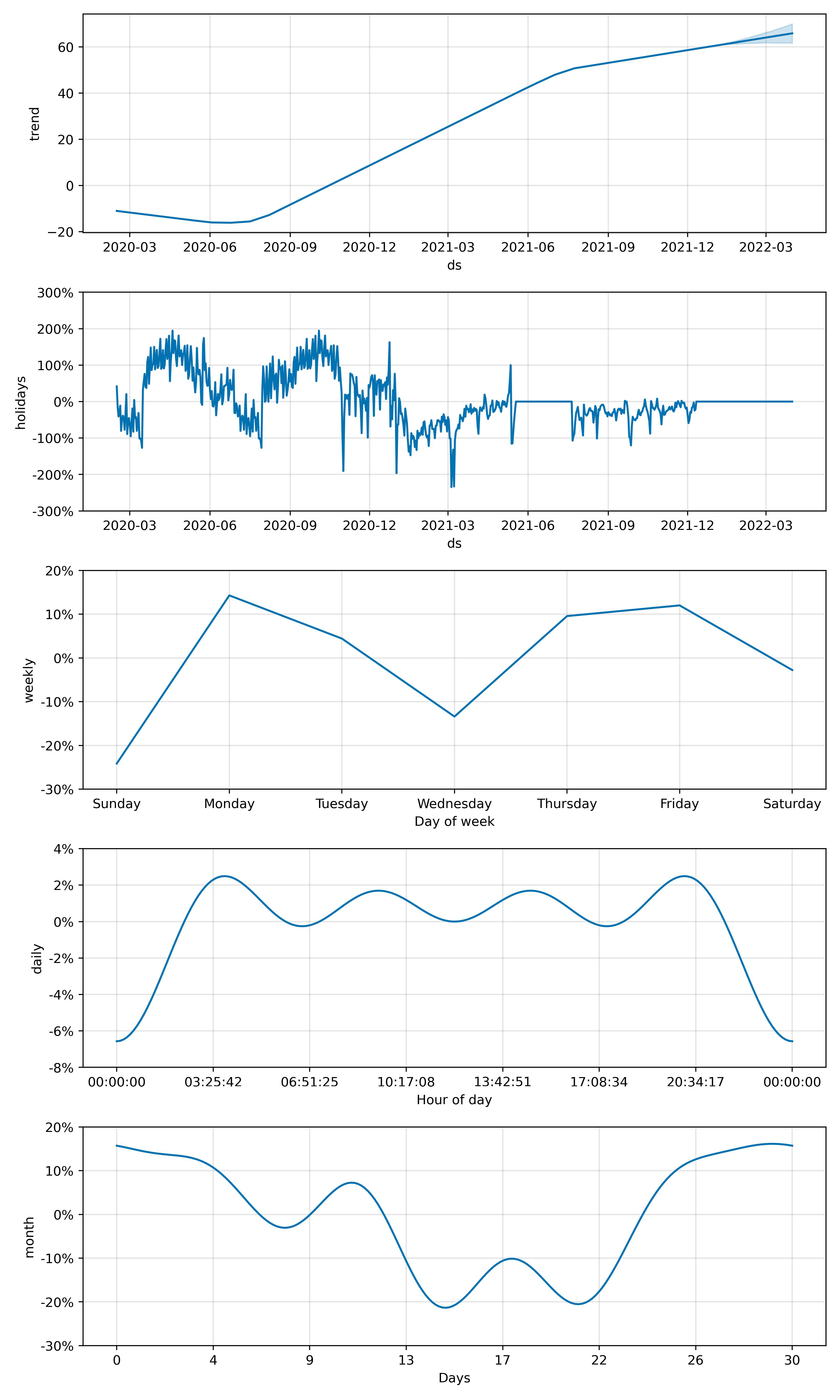}
\caption{The components of the improved model\label{comp_of_impr}}
 \end{figure}
 
 \noindent According to the improved Model, the trend dropped significantly in 2020 as a result of  the irregular fluctuations caused by the emergence of the covid-19 pandemic, and then began to rise as a result of the pandemic's eradication and a strong mobilization of the population.\\
The weekly traffic pattern illustrates the peak at the beginning of the week, probably in the coming quarter compared to the historical base. Considering the mobility behavior of the population, the population invaded the places of commercial activities, such as food stores, warehouses, drugstores, and pharmacies. This means that there will be a high demand for transportation and a high traffic intensity in the future of the historical area, which in all probability will lead to the presence of congestion on the road, hence the traffic. The majority of the population is at high risk of traffic on weekdays, but Saturday and Sunday are less risky. Most likely because of the fluctuations due to the nature of these days (weekend rest, churches), there is less movement of people on these days.\\
In the components, there is also a close view of the hours; this is explained by the peak hours, when the hours begin to move a lot in the interval of 03:00-06:00. A large number of people will start to leave their homes around these residences during these peak hours. \\

\noindent {\bf Measuring the performance of the improved model: cross-validation} \\
Prophet includes a time series cross-validation feature to measure forecast error using historical data.\\
%-%
\begin{table}[!h]%[!h]
{\small
\begin{center}
\caption{Cross-validation of the improved model forecasting\label{cross_vali}}
\begin{tabular}{ccccccc}
\hline
            & {\bf ds}  &  {\bf yhat} & {\bf yhat\_lower }  & {\bf yhat\_power} & {\bf y} & {\bf cuuoff} \\
\hline
{\bf 0}  &     2021-12-27	 & 	51.090203	&  40.850683     &  60.665832    & 53.0		&      2021-12-26  \\
\hline
{\bf 1}  &       2021-12-28	 & 	49.229138	&   39.913130    &     59.664335  & 	49.0	&     2021-12-26     \\
\hline
{\bf 2}  &    2021-12-29   	 & 	45.066784	&  34.353272     &   55.147226 	 & 	51.0	&      2021-12-26    \\
\hline
{\bf 3}  &    2021-12-30    	 & 	65.294401	&  55.154308     &  74.853545   & 	64.0	&       2021-12-26   \\
\hline
{\bf 4}  &       2021-12-31	 & 	69.364065    &    58.592437     &    78.890927   	 & 	66.0	&        2021-12-26  \\
\hline
\end{tabular} 
\end{center}
}
\end{table}

\noindent For the last day, the true value $yhat$ = 69.364065 and the predicted value $y$ = 66.0, which gives an error of about 5\%, according to the performance measure "mape", and "mdape".  
The detail is described in the performance measure table (see Table \ref{err_imp}) of the improved model.\\
%-%
\begin{table}[!h]%[!h]
{\small
\begin{center}
\caption{The errors of the improved model\label{err_imp}}
\begin{tabular}{ccccccccc}
\hline
 {\bf  }   & {\bf horizon}  &  {\bf mse} & {\bf  rmse}  & {\bf mae} & {\bf mape}  & {\bf mdape} & {\bf smape} & {\bf coverage} \\
\hline
{\bf 0}  &    1 days	 & 	3.647286	&   1.909787     &   1.909787   & 	0.036034	&   0.036034    & 	0.036695	&    1.0  \\
\hline
{\bf 1}  &    2 days 	 & 	0.052504	&   0.229138    &    0.229138    & 	 0.004676	&   0.004676      & 	0.004665	&   1.0   \\
\hline
{\bf 2}  &   3 days   	 & 	35.203048	&   5.933216    &   5.933216    	 & 	0.116338	&    0.116338    &   0.123523		&   1.0     \\
\hline
{\bf 3}  &   4 days  	 & 	1.675473	&   1.294401    &   1.294401   	 & 	0.020225	&    0.020225     & 	0.020023	&    1.0    \\
\hline
{\bf 4}  &  5 days    	 & 	11.316931	&   3.364065    &   3.364065  	 & 	0.050971	&     0.050971    & 	0.049704	&   1.0        \\

\hline
\end{tabular} 
\end{center}
}
\end{table}

%-%
\begin{figure}[!h]
 	\centering
\includegraphics[width=11.5cm,height=8.cm]{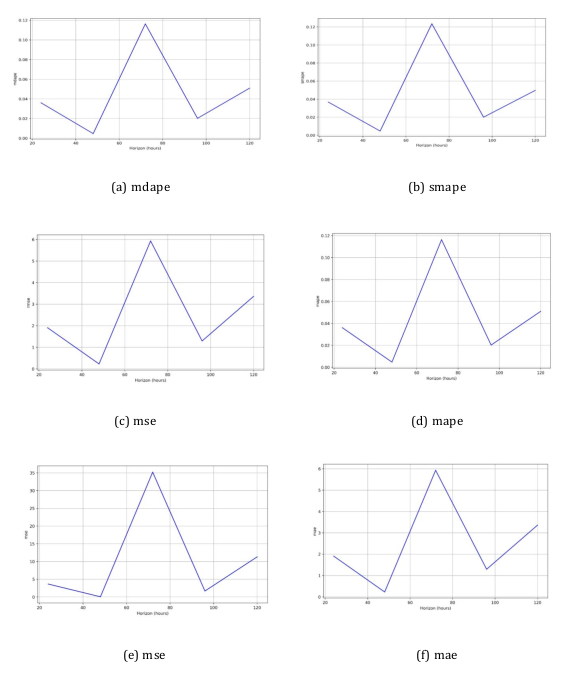}
\caption{The errors of the improved model\label{er_imp}}
 \end{figure}
 
\noindent Our different errors have a decreasing tendency, even if it seems to go up, but most of them represent a peak of 12\% of error. This explains why we can trust our model in an uncertainty interval of 95\%.

%--%
\section{Conclusion and future works}\label{ccl}
In this article, we were able to perform a traffic prediction task in the city of Dakar, Senegal, using mobility data. 
The developed algorithms to learn from data have a main objective: to orient the decisions of decision-makers in an optimal way to plan, set goals, and detect anomalies. They inform future information on past facts and also give a cross-sectional view of the business on the range of wise and orthonormal orientations for decision-makers within the framework of reflexive, useful, and optimal decision-making. \\
This study will allow decision-makers to control the actions, carry out appropriate policies, and reflect on the congestion caused by the mobility of the population observed on the timeline at monthly, weekly, and daily frequencies. 
Thus, decisions can be decided otherwise to constrain any maneuver of traffic congestion caused by population mobility. \\
In future works, we can implement a system of real-time information acquisition about people's mobility as well as traffic information, such as information on road traffic, the speed of vehicles on a stretch, the presence of accidents. With extended time intervals,  it is also possible to develop other methods for a possible comparative study.

%---%
\section*{Acknowledgement}
\addcontentsline{toc}{section}{Acknowledgement}
\noindent The authors thank the anonymous referees for useful comments and suggestions. Gratitude is expressed to the projects "Impulse Fund for the Scientific and Technical Research" (FIRST) and  "Non Linear Analysis, Geometry and Applications Project" (NLAGA) for the support of this work.

%---%
\bibliographystyle{plainnat}
\bibliography{biblio_ref}
			
\end{document}